\documentclass[mathleft]{an}
\usepackage{graphicx}
\usepackage{times}

\newcommand{\Om}{{\it \Omega}}

\def\lsim{\lower.4ex\hbox{$\;\buildrel <\over{\scriptstyle\sim}\;$}}
\def\beg{\begin{eqnarray}}
\def\ende{\end{eqnarray}}
\def\q{\qquad}
\renewcommand{\vec}[1]{\textbf{#1}}
\Pagespan{1}{}
\Yearpublication{2008}
\Yearsubmission{2008}
\Month{}
\Volume{329}
\Issue{}
\DOI{}
\begin{document}
\titlerunning{Tayler instability  and   Hall effect in PNS}
\authorrunning{G. R\"udiger et al.}

\title{The Tayler instability and  the Hall effect in protoneutron stars}

\author{G. R\"udiger\inst{1}\fnmsep\thanks{Corresponding author: gruediger@aip.de} \and M. Schultz\inst{1} \and M. Mond\inst{2} \and D.A. Shalybkov\inst{3}}
\institute{Astrophysikalisches Institut Potsdam, An der Sternwarte 16,
D-14482 Potsdam, Germany
\and
Department of Mechanical Engineering, Ben-Gurion University of the Negev, P.O. Box 653, Beer-Sheva 84105, Israel
\and
A.F. Ioffe Institute for Physics and Technology, 194021, St. Petersburg, Russia}

\received{2008} \accepted{2008} \publonline{2008}
\keywords{neutron stars -- instabilities -- magnetohydrodynamics -- magnetic fields -- plasmas}

\abstract{%
Collapse calculations indicate that the hot newly born protoneutron stars (PNS) rotate
differentially so that strong toroidal magnetic field components should exist in the outer crust where   also  the Hall effect appears to be important when the Hall parameter $\hat\beta=\omega_B \tau$  is of order unity.
The  amplitudes of the induced toroidal magnetic fields are limited from above by the Tayler  instability. An important  characteristic of the Hall effect  is its distinct dependence on the {\em sign} of the magnetic field. We find for fast rotation that positive (negative) Hall parameters essentially  reduce (increase) the stability domain. It is thus concluded that the toroidal field belts in PNS induced by their differential rotation should have different amplitudes in both  hemispheres which later are frozen in. Due to the effect of magnetic  suppression of the heat conductivity also the brightness of the two hemispheres should be different. As a possible example for our scenario the  isolated neutron star RBS~1223 is considered which has been found to exhibit different X-ray brightness at both hemispheres.
}

\maketitle
\section{Introduction}

Progenitors of neutron stars are high-mass stars with more than eight solar masses
that develop  a degenerate iron core. If the core mass approaches  the
Chandrasekhar limit it becomes gravitationally  unstable  and implodes.
The collapse comes  to a temporary end if nuclear densities are
reached. At that stage  the rebounding inner core drives a shock wave into the
outer core, a mechanism that is currently believed to be responsible for the appearance of supernova.

If the
 core of the supergiant rotates already rapidly  the neutron star will be
born as a fast rotator with an angular velocity near the break-off
 value, i.e.\ 1 kHz. This value exceeds the rotation  rate of the
fastest young pulsars known
by one  order of magnitude so that  the question arises how  a critically
rotating protoneutron star (PNS) spins down. One  possibility  is the
angular momentum loss by gravitational wave emission via unstable r-modes (Friedman \& Schutz
1978; Andersson 1998; Stergioulas \& Font 2001; Lindblom, Tohline \&  Vallisneri
2001).  As the viscous damping of the r-modes is smallest at  temperatures
around $10^9$ K, this  instability  works best as long as the neutron  star remains
hot. Up to 90\% of the rotational energy can be removed in that  way from
the newly formed neutron star within hours. Other possibilities involve  angular
momentum transport due to nonaxisymmetric instabilities also connected with   gravitational
waves.

Following  Burrows (1987), entropy-driven
convection may play an essential role in the neutrino-mediated supernova  explosion
scenario since it enhances the neutrino
luminosities in the  post-collapse stage. Such a
convection might be   important with regard to a standard-dynamo action in PNS (Thompson  \& Duncan 1993).
 If, on the other hand, additionally differential rotation exists in the turbulent domain, then an $\alpha\Om$-dynamo can work producing strong toroidal magnetic fields (Bonanno et al. 2003, 2006).
Indeed, hydrodynamic simulations of  rotational
supernova core collapse  have shown that even a nearly  rigidly
rotating initial core results in a strongly
differentially rotating  post-collapse neutron star (M\"onchmeyer \& M\"uller 1989; Janka \& M\"onchmeyer 1989; Dimmelmeier, Font \&
M\"uller 2002; Kotake et al.
2004; Ardeljan et al. 2005; Burrows et al. 2007). Any nonhomologous collapse
creates necessarily some degree  of differential rotation if angular momentum is
conserved locally during  collapse. The  models reveal a strong
differential rotation in the  azimuthally averaged angular velocity (Ott et al.  2005).  Differential rotation may furthermore be generated by
r-modes via nonlinear  effects (Rezzolla, Lamb \& Shapiro 2000) or simply by
accreting falling-back  material (Watts \& Andersson 2002).
\subsection{Differential rotation and magnetic fields}
In the presence of a poloidal field $B_R$, differential rotation with a shear $q$  produces a toroidal field component by induction. The ratio of the resulting field to the original one can  simply be estimated as
\beg
\epsilon\equiv \frac{B_\phi}{B_{R}}\simeq q\Om \, \tau \quad {\rm if} \quad \epsilon< {\rm Rm},
\label{tor}
\ende
with Rm as the magnetic Reynolds number of the differential rotation. For  high Rm the
differential rotation  may induce strong toroidal fields. Also flux compression will play an important role in amplification of both poloidal fields and toroidal fields (Burrows et al. 2007).  However, the resulting magnetic field transports angular momentum outwards and feedbacks the differential rotation. The timescale of this backreaction is
\beg
\tau=\frac{\mu_0\, \rho \, q\Om L^2}{B_R B_\phi}= \frac{\mu_0\, \rho \, q\Om L^2}{\epsilon B_R^2}.
\label{taul}
\ende
With $\epsilon \simeq q\Om\tau$ one finds
\beg
\tau\simeq \frac{\sqrt{\mu_0 \rho} L}{B_R},
\label{tausim}
\ende
i.e. the Alfv\'en travel time of  $ 1\dots 10$ s (Shapiro 2000). Hence, after (\ref{tor}) $\epsilon \simeq 100\dots 1000$.
Typical values of the neutron stars have been used: $\rho\simeq 10^{13}$ g/cm$^3$, $\Om\simeq 100$ s$^{-1}$, $L\simeq 10^5$ cm (the crust thickness)  and $B_r\simeq 10^{12}$ G. Note that for $\epsilon\simeq 1000$ the differential rotation is immediately destroyed. We find $\epsilon < 1000$ (i.e. $B_{\rm tor} < 10^{15}$ G) as a necessary condition for the existence of differential rotation over several rotation periods. This value is a rather small value insofar
as 
\beg
\epsilon_{\rm max}\simeq {\rm Rm}= \frac{\Om L^2}{\eta}\simeq \frac{10^{12}\ {\rm cm^2/s}}{\eta},
\label{eps}
\ende
so that already for $\eta<10^{9}$ cm$^2$/s the critical $\epsilon$ is exceeded. The microscopic value is only $\eta\simeq 10^{-6}$ cm$^2$/s. It is obvious that with such a small microscopic value a differential rotation cannot survive. It is  an open question whether such high values of $\eta$ can be reached in neutron stars (see Naso et al. 2007).

The diffusion time $L^2/\eta$ for $\eta \simeq 10^{9}$ cm$^2$/s is also  10 s which, however, would also be the decay time of the differential rotation for magnetic Prandtl number ${\rm Pm}\geq 1$. With such high values of viscosity a prescribed differential rotation cannot exist longer than a few rotations.

In the present paper we assume that differential rotation exists for at least 10 s ($\simeq 100$ rotations). During this time r-modes are excited  producing gravitational waves. The viscosity must thus be $\lsim 10^9$ cm$^2$/s. This value corresponds to the expression $\nu\simeq \alpha L^2 \Om$ with $\alpha\simeq 10^{-3}$ which is known from the accretion theory (due to small-scale MRI). As the microscopic magnetic Prandtl number Pm is very large for neutron stars we also continue with a high magnetic Prandtl number for the unstable PNS crust matter, say ${\rm Pm}\simeq 100$ so that $\eta\simeq 10^7$ cm$^2$/s. In that case the magnetic diffusion time exceeds the viscous time by a factor of 100 and the magnetic Reynolds number is of order $10^5$. The differential rotation would thus generate huge toroidal fields with $\epsilon\simeq 10^5$ which, however,  would destroy the  differential rotation.

Therefore, in the present paper the stability of strong toroidal magnetic  fields  against nonaxisymmetric perturbations  is probed in order to find their real upper limits.  We are thus considering the Tayler instability under the influence of differential rotation and for high magnetic Prandtl numbers. The toroidal field is assumed to dominate  the poloidal field ($\epsilon>1$) so that stability under only toroidal field  is considered. To that end, as will be shown in the next section, also the influence of the Hall effect in neutron stars must be taken into account.

\subsection{Magnetic fields and Hall effect in neutron stars}
Neutron stars have the strongest magnetic fields found in the Universe, with fields exceeding $10^{13}$ G for young ($\sim 10^7$ yr) radio and X-ray pulsars, and a still appreciable $10^8$--$10^{10}$ G for much older ($\sim 10^{10}$ yr) millisecond pulsars. This correlation between field strength and age suggests that these very different strengths are due to the field decaying in time rather than to  differences between different neutron stars.

Jones (1988) and Goldreich \& Reisenegger (1992) have proposed that correlation between the magnetic field of the neutron star and its age is due to the Hall drift. Since the Hall effect enters the evolution equation for $\vec{B}$ as a quadratic nonlinearity, it necessarily leads to a timescale inversely proportional to $|\vec{B}|$.  The Hall effect is therefore attractive  for explaining the  variations in the decay rates (for $B\sim 10^{13}$ G the field should evolve on a $10^7$ year timescale while if $B\sim 10^{10}$ G it should evolve on a $10^{10}$ year timescale).

There is a bulk of literature about the existence  of the Hall effect in neutron stars. The main findings may be summarized as follows. The Hall effect strongly depends on the magnetic field amplitude and the temperature of the neutron star.
In the presence of strong magnetic fields the magnetic diffusivity
is anisotropic and is given by a tensor whose components
along the magnetic field are $\eta_\parallel$, the components perpendicular
to the magnetic field are $\eta_\perp$, and off-diagonal Hall
component $\eta_{\rm H}$. For more details
concerning the generalized Ohm's law in multi-component plasma
we refer to the papers Yakovlev \& Shalybkov (1991) and
Shalybkov \& Urpin (1995).

With Hall effect included the magnetic induction equation
takes the general form
\beg
\frac{\partial \vec{B}}{\partial t}- \eta \Delta \vec{B}= {\rm curl}(\vec{u}\times \vec{B} - \beta\, {\rm curl}\vec{B}\times \vec{B})
\label{1a}
\ende
with $\eta \equiv \eta_\perp$ and $\beta =c/4\pi e n_e$, where $n_e$ is the electrons' number density. In addition, it is useful to define the Hall magnetic diffusivity as   $\eta_{\rm H} \equiv  \beta B$.
The Hall effect becomes important if $\hat\beta > 1$, where
\beg
\hat\beta= \frac{\eta_{\rm H}}{\eta_\perp}.
\label{1b}
\ende
For magnetic fields smaller than some critical value,
$B_{\rm cr}$, $\eta_\perp=\eta_\parallel=\eta_0$ where $\eta_0$ is the magnetic
diffusivity without an applied magnetic field. If $B>B_{\rm cr}$ than $\eta_{\perp}$ increases as
$B^2$ for increasing magnetic field. The Hall magnetic diffusivity, on the other hand,
is proportional to the magnetic field value. As a result, the Hall effect can be important
only in some vicinity of the $B_{\rm cr}$.

The critical magnetic field can vary significantly
within the neutron star envelopes depending on chemical composition,
temperature and density. According to Potekhin (1999) the critical magnetic
field is $\sim 10^{12}$G for iron composition with temperature $10^8$ K and
density $10^{11}$ g/cm$^3$. Detailed calculations of the electrical conductivity in pure neutron star crusts (Cumming et al. 2004) indicate that the Hall time scale under such parameters is indeed shorter than the Ohmic decay time, which means that $\hat \beta > 1$.   One finds\footnote{see http://www.ioffe.rssi.ru/astro/conduct/} that for iron (Z=26, A=56)  with $\rho=10^{13}$ g/cm$^3$ the $\hat\beta$ varies from $10^{-3} B_{12}$ for $T=10^{10}$ K to $3 B_{12}$ for $10^8$ K.  Note, however, that  for  the same plasma  the Hall
parameter $\hat\beta$ reaches a maximal value of
$\sim 10$ for $B \sim 10^{13}$ G and decreases for higher magnetic
field values.

Hence, it makes sense to ask for the consequences of the Hall term for young neutron stars  with fields of $B_{12}\gg 1$ which can be imagined -- and this is the point here -- as toroidal field due to the induction of a differential rotation.
Important for us is only the assumption   that the (early) phase of the existence of differential rotation in the crust of the PNS is accompanied by $\hat\beta$ of order unity for the resulting {\em toroidal} fields. Obviously, $\hat\beta$ linearly depends on the magnetic amplitude so that  we can write
 \beg
\hat\beta= \beta_0\ {\rm \sqrt{Pm} Ha}=\beta_0\ {\rm S},
\label{hb}
\ende
with S as the Lundquist number (see below).
The parameter $\beta_0$ does not depend on the magnetic field\footnote{Other possible notations for $\hat\beta$ are $\hat\beta={\rm Rb}=a_e=\omega_B\tau$}.

To estimate the magnetic Prandtl number we should
also use $\eta_{\rm \perp}$ instead of $\eta_0$. We will have smaller magnetic
Prandtl numbers for the parameters where the Hall effect is important. Nevertheless,
it is easy to estimate that the magnetic Prandtl number can be much larger than 1
for the typical  neutron star envelopes parameters.

\section{Tayler instability}
Differential rotation leads to an increase of the toroidal component of the magnetic  field  by winding up the
poloidal field lines.  The ratio of both components is given by the magnetic Reynolds number of the
differential rotation which is very large also for PNS due to their small values of the magnetic diffusivity $\eta$.
The same is  true for the microscopic magnetic Prandtl
number ($\nu\simeq 10$ cm$^2$/s, $\eta\simeq 10^{-6}$ cm$^2$/s).

Too strong toroidal fields, however,  become unstable against the Tayler instability.
Tayler (1961, 1973) and Vandakurov (1972)  considered the stability of nonaxisymmetric disturbances  and
showed that for an ideal fluid the necessary and sufficient condition
for stability is
\beg
 \frac{\rm d}{{\rm d}R}( R B_\phi^2) < 0.
\label{tay}
\ende
An almost uniform field would therefore be  stable against axisymmetric perturbations but
 unstable against nonaxisymmetric perturbations (with $m=1$ being the most unstable mode).
Differentially rotating PNS might be susceptible to that kink-type instability that in turn
may limit their magnetic
field amplitude.

\begin{figure}[htb]
\includegraphics[width=8.0cm,height=5.5cm]{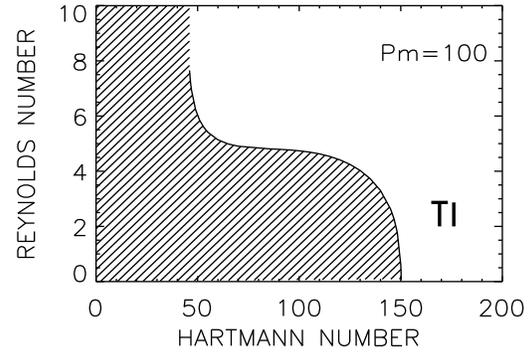}
\caption{The stability domain  (hatched) for outer cylinder  rotating with 50\% of the inner cylinder ($\mu_\Omega=0.5$), the magnetic field is almost uniform ($\mu_B=1$),  the perturbations are nonaxisymmetric ($m=1$). Note the destabilizing action of high magnetic Prandtl numbers (here $\rm Pm=100$).
\label{Tay}}
\end{figure}

Criterion (\ref{tay}) cannot be applied directly to fields under the influence of differential rotation.
In a first step to understand the complicated interaction
of toroidal magnetic fields and differential rotation we have studied
a Taylor-Couette container with two corotating cylinders where the radial
rotation law is hydrodynamically stable.  An electric  current
is flowing parallel to the rotation axis through the conducting fluid, thus producing a nearly uniform toroidal magnetic field.  It
becomes unstable against nonaxisymmetric perturbations for non rotating
cylinders but only for a rather strong magnetic field. If measured in terms
of Hartmann numbers,
\beg
{\rm Ha} = \frac{B_0 R_0}{\sqrt{\mu_0\rho\nu\eta}},
\label{Ha}
\ende
with $R_0=\sqrt{(R_{\rm out}-R_{\rm in})R_{\rm out}}$, this is
at about ${\rm Ha} = 150$ (see Fig.~\ref{Tay}). In case of rotating
cylinders without magnetic field the rotation law may be so flat that it is hydrodynamically stable.
With magnetic fields it becomes always unstable (R\"udiger et al.  2007). One finds that differential
rotation is strongly {\em destabilizing for large magnetic Prandtl numbers} which are characteristic for  the PNS
matter. One can thus expect
that the toroidal magnetic fields  induced by the differential rotation of PNS
are  limited by the described  current-induced instability.

In the present paper the  stability problem for strong toroidal magnetic fields under the influence of  differential rotation and the Hall effect is  considered. We shall see that in this case even the sign of the toroidal field (with respect to the global rotation) will play an important role.
\section{The basic equations}

\begin{figure}[htb]
\includegraphics[scale=0.45]{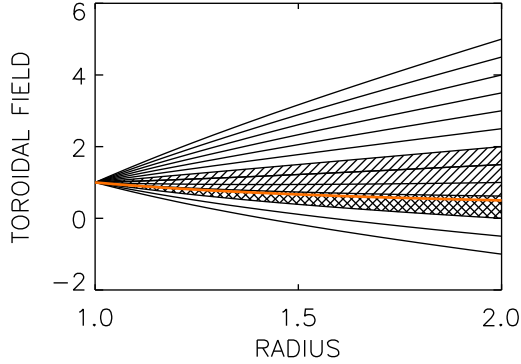}
\caption{The possible radial profiles of the toroidal magnetic field between the two cylinders. The value of the intersection of each of the profiles with the right vertical axis is the corresponding $\mu_B$-value. The profiles in the hatched domain are stable against axisymmetric perturbations while the cross-hatched area is also stable against nonaxisymmetric perturbations. The current-free solution $B_\phi\propto 1/R$ is given by the red line.}
\label{fig1}
\end{figure}

The  basic state in the cylindrical system is
$
U_R=U_z=B_R=B_z=0$
and
\beg
B_\phi=A R+\frac{B}{R},  \q
U_\phi=R\Om=a R+\frac{b}{R},
\label{basic}
\ende
where $a$, $b$, $A$ and $B$ are constant values defined by
\beg
a=\Om_{\rm{in}}\frac{\mu_\Omega-{\hat\eta}^2}{1-{\hat\eta}^2}, \q
b=\Om_{\rm{in}} R_{\rm{in}}^2 \frac{1-\mu_\Omega}{1-{\hat\eta}^2},
\nonumber \\
 A=\frac{B_{\rm{in}}}{R _{\rm{in}}}\frac{\hat \eta
(\mu_B - \hat \eta)}{1- \hat \eta^2},  \ \ \
B=B_{\rm{in}}R _{\rm{in}}\frac{1-\mu_B \hat\eta}
{1-\hat \eta^2}.
\label{ab}
\ende
Here  is
\beg
\hat\eta=\frac{R_{\rm{in}}}{R_{\rm{out}}}, \; \; \;
\mu_\Omega=\frac{\Om_{\rm{out}}}{\Om_{\rm{in}}},  \; \; \;
\mu_B=\frac{B_{\rm{out}}}{B_{\rm{in}}},
\label{mu}
\ende
with $R_{\rm{in}}$ and $R_{\rm{out}}$ as the radii,
$\Om_{\rm{in}}$ and $\Om_{\rm{out}}$  the angular velocities,
and $B_{\rm{in}}$ and $B_{\rm{out}}$
as the azimuthal magnetic fields of the inner and the outer cylinders. The possible magnetic field solutions which do not decay  are plotted in  Fig. \ref{fig1}.

We are interested now in the linear stability of the background state  (\ref{basic}).
In that case, the perturbed  quantities of the system  are given  by
\beg
u_R, \; R\Om+u_\phi , \; u_z, \; b_R,  \; B_\phi+b_\phi, \; b_z.
\label{pert}
\ende
As usual, the perturbations are developed in normal modes of the form

\beg
F=F(R){\rm{exp}}({\rm{i}}(kz+m\phi+\omega t)).
\label{nmode}
\ende

Terms of the form (\ref{pert}) and (\ref{nmode}) are inserted now  into the the induction equation (\ref{1a}) with the Hall effect included and linearized about the background state. The result is:
\beg
\frac{\partial \vec{B}}{\partial t}- \eta \Delta \vec{B}= \vec{E}-\beta \vec{H}
\label{1}
\ende
with
\beg
E_R=\frac{1}{R} \left({\rm i} m u_R B_\phi-{\rm i} m R \Om b_R\right),
\label{2}
\ende
\beg
\lefteqn{E_\phi=- \frac{{\rm d} B_\phi}{{\rm d} R} u_R +\Om R \frac{{\rm d} b_R}{{\rm d} R}  + \frac{{\rm d} \Om}{{\rm d} R}R b_R  - }\nonumber\\
&& \quad -B_\phi\frac{{\rm d} u_R}{{\rm d} R}  - B_\phi {\rm i} k u_z +\Om b_R  +  {\rm i} k \Om R b_z,
\label{3}
\ende
\beg
E_z= \frac{1}{R} \left({\rm i} m u_z B_\phi-{\rm i} m R \Om b_z\right)
\label{4}
\ende
and the Hall terms
\beg
\lefteqn{H_R= \frac{1}{R^2} \bigg(-\frac{{\rm d} B_\phi}{{\rm d} R} {\rm i} k R^2 b_R +
 B_\phi  k m R b_\phi-}\nonumber\\
&& \quad - B_\phi  {\rm i} k R b_R- B_\phi  m^2 b_z\bigg),
\label{5}
\ende
\beg
\lefteqn{H_\phi= \frac{1}{R^2}\bigg(-\frac{{\rm d} B_\phi}{{\rm d} R}  {\rm i} m R b_z-  B_\phi {\rm i} m R \frac{{\rm d} b_z}{{\rm d} R}-}\nonumber\\
&& \quad - 2 B_\phi  {\rm i} k R b_\phi-
B_\phi  k m R b_R+ B_\phi  {\rm i} m b_z\bigg),
\label{6}
\ende
\beg
\lefteqn{H_z= \frac{1}{R^2}\bigg(\frac{{\rm d} B_\phi}{{\rm d} R^2}  R^2 b_R+ \frac{{\rm d} B_\phi}{{\rm d} R} R^2\frac{{\rm d} b_R}{{\rm d} R}  + \frac{{\rm d} B_\phi}{{\rm d} R}  {\rm i} m R b_\phi+}\nonumber\\
&& \quad\quad + 2\frac{{\rm d} B_\phi}{{\rm d} R}R b_R  +  B_\phi {\rm i} m R \frac{{\rm d} b_\phi}{{\rm d} R}
 + B_\phi R \frac{\partial b_R}{\partial R}  +\nonumber\\
&& \quad\quad\quad\quad +2 B_\phi  {\rm i} m b_\phi+ B_\phi  m^2 b_R\bigg).
\label{7}
\ende

The dimensionless numbers of the problem are the magnetic Prandtl number (Pm), the
Hartmann number (Ha) and the Reynolds number (Re), i.e.
\beg
{\rm{Pm}} = \frac{\nu}{\eta}, \ \ \
{\rm{Ha}}=\frac{B_{\rm{in}} R_0}{\sqrt{\mu_0 \rho \nu \eta}},  \ \ \
{\rm{Re}}=\frac{\Om_{\rm{in}} R_0^2}{\nu},
\label{pm}
\ende
where $R_0=(R_{\rm{in}}(R_{\rm{out}}-R_{\rm{in}}))^{1/2}$
is the characteristic length scale, $\nu$  the kinematic viscosity and $\eta$  the
magnetic diffusivity. The magnetic Reynolds number is ${\rm Rm =Pm\ Re}$ and the Lundquist number is ${\rm S =\sqrt{Pm}\ Ha}$.

We use $R_0$ as a unit of length and $R_0^{-1}$ as a unit of the wave number,
$\eta/R_0$ as a unit of the perturbed velocity, $\Om_{\rm in}$
as a unit of angular velocity and $\omega$, and $B_{\rm in}$ as a unit of
magnetic fields (basic and disturbed).

In normalized quantities eq. (\ref{1}) may be cast in the following form:
\beg
{\rm i} \omega {\rm Rm} \ \vec{b} = D(\vec{b}) + \hat{\vec{E}} - \hat\beta \vec{H}
\label{8}
\ende
with
\beg
\hat E_R= \frac{1}{R}\left({\rm i} m \hat B u_R -{\rm i} m R\, {\rm Rm}\ \hat\Om b_R\right),
\label{10}
\ende
\beg
\lefteqn{\hat E_\phi= - \hat B' u_R - {\rm i} k \hat B u_z - \hat B \frac{{\rm d}u_R}{{\rm d}R}+{\rm Rm}}\nonumber\\
&&   \bigg(R \hat\Om\frac{{\rm d} b_R}{\partial R}  + R\frac{{\rm d} \hat \Om}{{\rm d} R} b_R + \hat\Om b_R  +  {\rm i} k  R \hat\Om b_z  \bigg),
\label{11}
\ende
\beg
\hat E_z= \frac{{\rm i}m}{R} \left(\hat B u_z -{\rm Rm}\hat\Om R b_z\right).
\label{121}
\ende
Here we have used the notations
\beg
\Om = \Om_{\rm in} \hat\Om
\quad {\rm and} \quad
B_\phi = B_{\rm in} \hat B.
\label{123}
\ende
The diffusion terms  are
\beg
\lefteqn{D_R(\vec{b})= \frac{{\rm d}^2 b_R}{{\rm d} R^2}- \frac{m^2}{R^2} b_R - k^2 b_R +}\nonumber\\
&& \quad\quad + \frac{1}{R} \frac{{\rm d} b_R}{{\rm d} R} - \frac{2{\rm i}m}{R^2} b_\phi -
 \frac{b_R}{R^2},
\label{13}
\ende
\beg
\lefteqn{D_\phi(\vec{b})= \frac{{\rm d}^2 b_\phi}{{\rm d} R^2}- \frac{m^2}{R^2} b_\phi - k^2 b_\phi +}\nonumber\\
&& \quad\quad + \frac{1}{R} \frac{{\rm d} b_\phi}{{\rm d} R} +\frac{2{\rm i}m}{R^2} b_R
 - \frac{b_\phi}{R^2}
\label{14}
\ende
and
\beg
D_z(\vec{b})= \frac{{\rm d}^2 b_z}{{\rm d} R^2}- \frac{m^2}{R^2} b_z - k^2 b_z + \frac{1}{R} \frac{{\rm d} b_z}{{\rm d} R}.
\label{15}
\ende
In the same way the normalized momentum equation can be written as
\beg
\lefteqn{{\rm Re}\left[\frac{{\partial } \vec{u}}{{\partial} t}+ \left(\vec{U}\nabla\right)\vec{u} + \left(\vec{u}\nabla\right)\vec{U}\right]=}\nonumber\\
&& \vec{D}(\vec{u})-\nabla P+
 {\rm Ha}^2\left({\rm curl} \vec{B} \times \vec{b} + {\rm curl} \vec{b}\times \vec{B}\right),
\label{16}
\ende
so that
\beg
{\rm i}\omega \, {\rm Re} \vec{u} + {\rm Re}\, \vec{G}= \vec{D}(\vec{u})-\nabla P + {\rm Ha}^2 \vec{L}
\label{17}
\ende
with
\beg
G_R= {\rm i} m \hat\Om u_R - 2 \hat\Om u_\phi,
\label{18}
\ende
\beg
G_\phi= (R^2 \hat\Om)' \frac{u_R}{R} + {\rm i} m \hat\Om  u_\phi,
\label{19}
\ende
\beg
G_z=  \hat\Om {\rm i} m  u_z
\label{20}
\ende
and
\beg
L_R= + \frac{{\rm i} m}{R} \hat B b_R - 2 \frac{\hat B}{R} b_\phi,
\label{21}
\ende
\beg
L_\phi= \frac{1}{R} (R \hat B)' b_R + \frac{{\rm i} m}{R} \hat B b_\phi,
\label{22}
\ende
\beg
L_z=  + {\rm i} \frac{m}{R} \hat B b_z.
\label{23}
\ende
The perturbed flow as well as the perturbed magnetic field are source-free, i.e.
\beg
\frac{{\rm d}u_R}{{\rm d}R}+\frac{u_R}{R}+{\textrm i}\frac{m}{R}u_\phi+{\textrm i}ku_z=0
\label{divu}
\ende
and
\beg
\frac{{\rm d}b_R}{{\rm d}R}+\frac{b_R}{R}+{\textrm i}\frac{m}{R}b_\phi+{\textrm i}kb_z=0.
\label{divb}
\ende
An appropriate set of ten boundary conditions is needed to solve  the system. No-slip conditions as well as zero normal components for the velocity on the walls
results in
\beg
u_R=u_\phi=u_z=0.
\label{ubnd}
\ende
The boundary conditions for the magnetic field depend on the electrical
properties of the walls. The tangential currents and the radial component
of the magnetic field vanish on conducting walls hence
\beg
\frac{{\rm d}b_\phi}{{\rm d}R} + \frac{b_\phi}{R} = b_R = 0.
\label{bcond}
\ende
These boundary conditions hold  both for $R=R_{\rm in}$ and for
$R=R_{\rm out}$.
\section{Results}

\begin{figure}[h]
\vbox{
\includegraphics[scale=0.45]{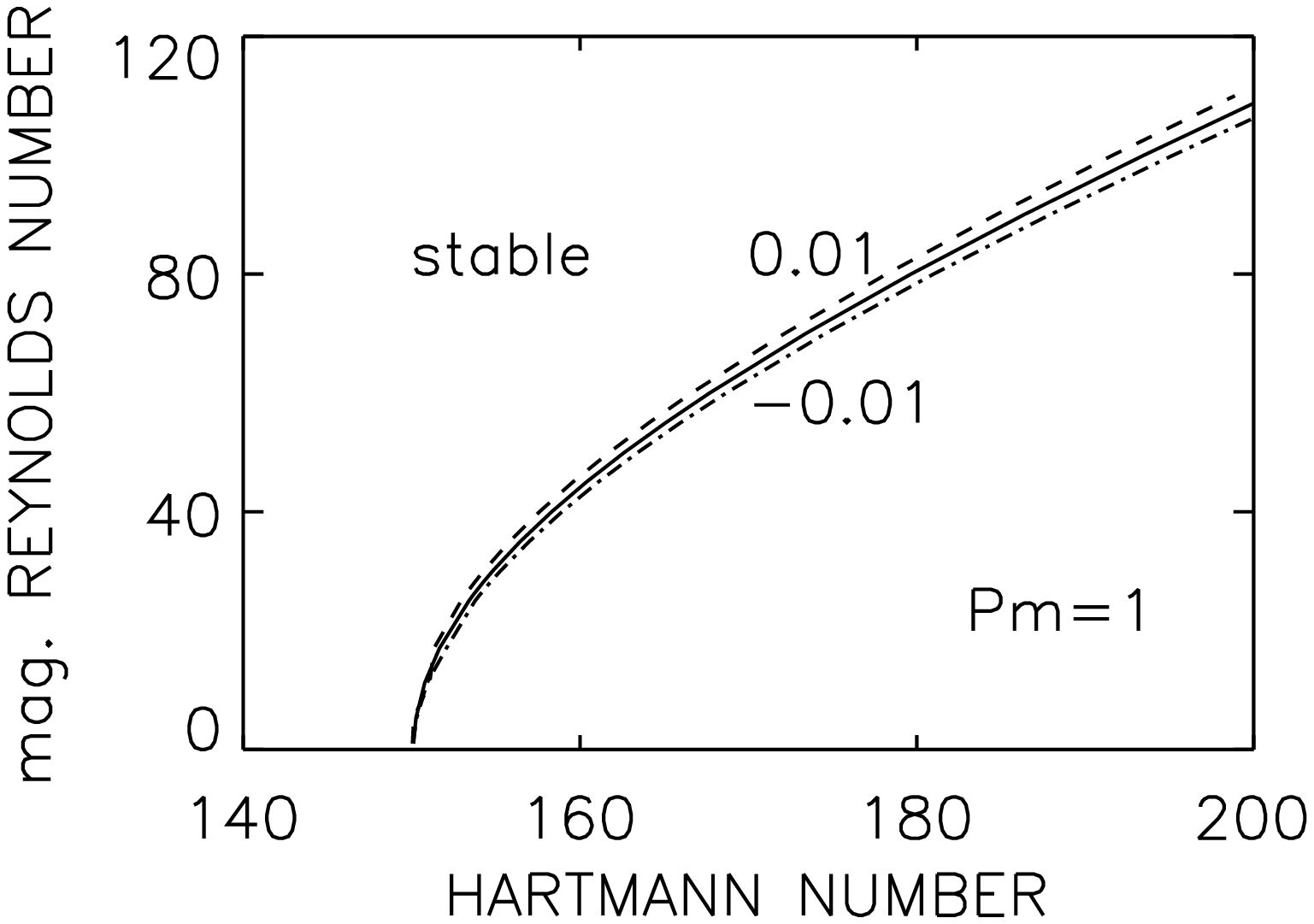}
\includegraphics[scale=0.45]{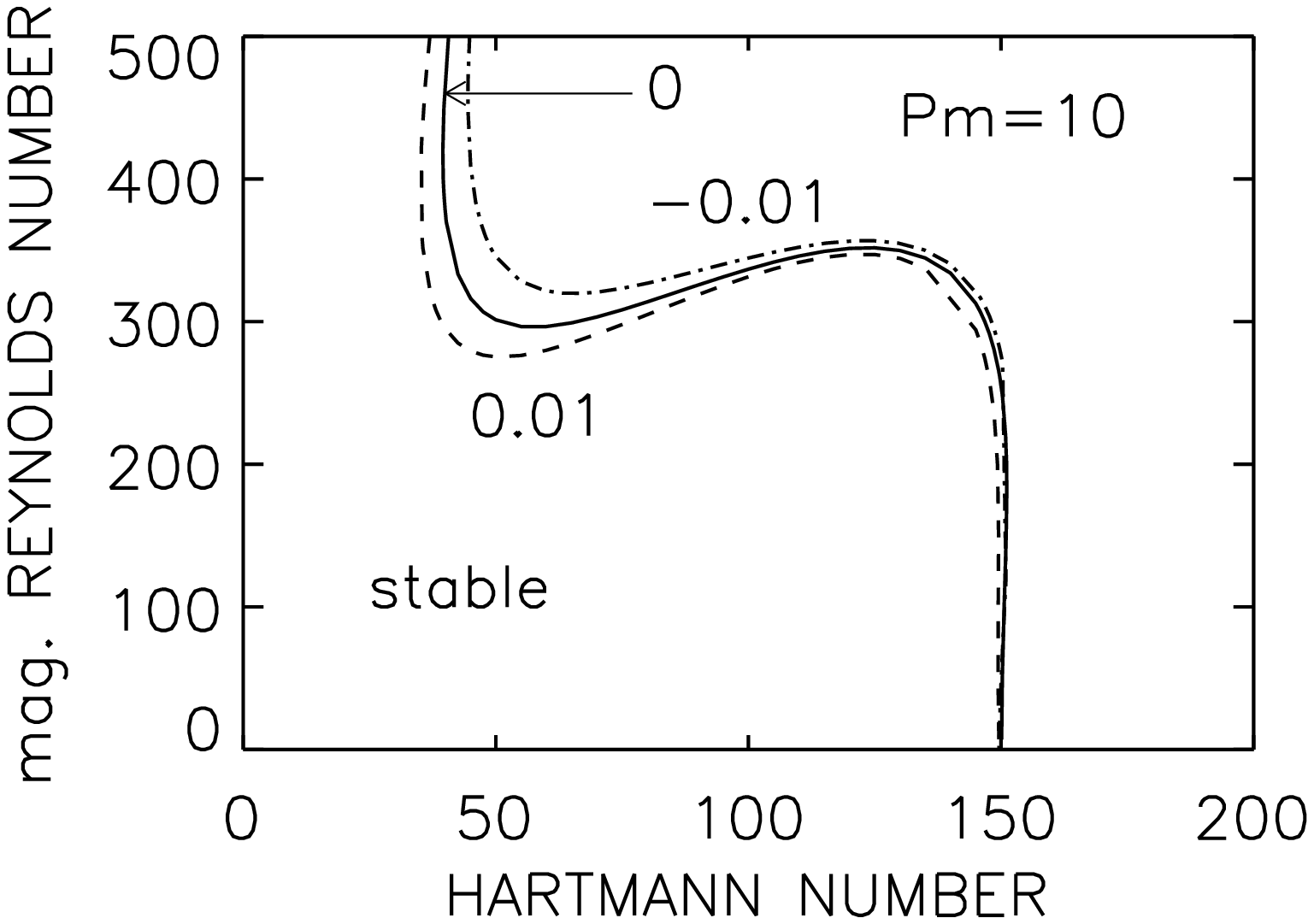}
\includegraphics[scale=0.45]{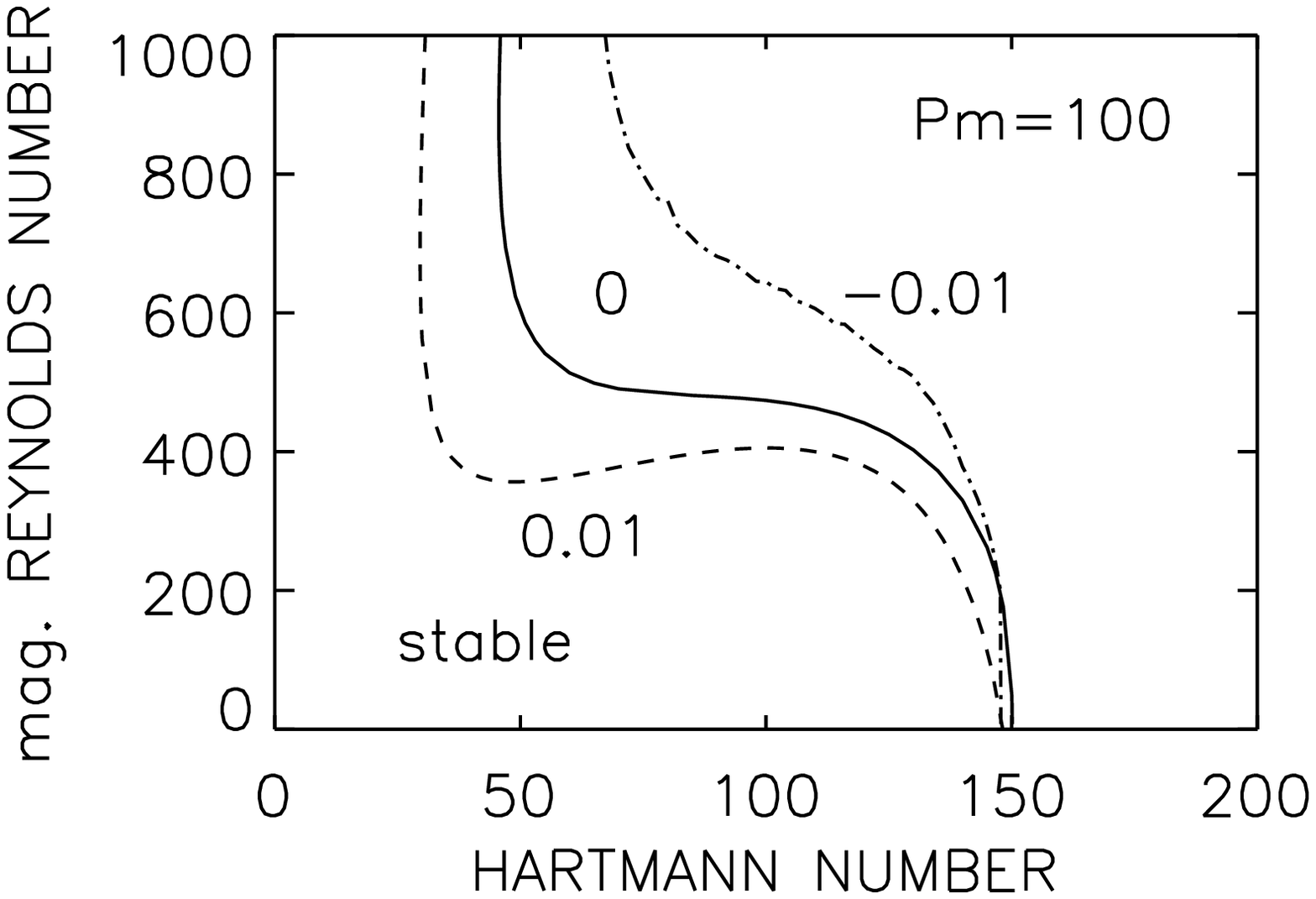}
}
\caption{Tayler instability ($m=1$) for various magnetic Prandtl numbers Pm and with Hall effect. $\mu_B=1$, $\mu_\Omega=0.5$. The curves are labeled by the Hall parameter $\beta_0$. Note that positive $\beta$ reduces the stability domain while negative  $\beta$ increases it.}
\label{mainplots}
\end{figure}

The equations have been solved for a simple model. The normalized gap width between the cylinders is 0.5 and the rotation law is rather flat approaching $\Omega\propto R^{-1}$ hence  $\mu_\Omega=0.5$. The toroidal field in the gap is almost uniform ($\mu_B=1$) but it is not current-free.  This field violates   (\ref{tay}) and is therefore Tayler-unstable with a  critical Hartmann number of about 150 (R\"udiger et al. 2007b).
This instability is strongly modified by the differential rotation. The results are given in the Fig. \ref{mainplots} for the Hall parameters $\beta_0= -0.01, 0 \ {\rm and} \  0.01$. The Hall parameter $\beta_0$ and the magnetic Prandtl number are the free parameters of the system. Note,
however, that due to (\ref{hb}) only $\sqrt{\rm Pm} \beta_0$ is a physical parameter in the definition of the
Hall quantity $\hat\beta$.
As  only the combination of $\beta_0 {\rm Ha}$ comes into the equations we can fix
the sign of $\beta_0$ and make the calculations for positive and negative Ha values
 or we can fix Ha as positive and
use both signs of $\beta_0$. We prefer the second possibility
so that the results for positive and negative $\beta_0$  correspond to  opposite
magnetic field orientations.

The solid line in Fig. \ref{mainplots} (bottom)  is identical with the marginal limit between stability and instability in Fig. \ref{Tay}. We find that the system is destabilized by the rotation for high magnetic Prandtl numbers ($\rm Pm=100$). In contrast, for $\rm Pm=1$ the rotation {\em stabilizes} the flow (Fig. \ref{mainplots}, top), which demonstrates very clearly the significant differences between the solutions with large and small magnetic Prandtl numbers.

In all these cases, however, the Hall effect acts in the same direction. For positive $\beta$ the stability domain is reduced and for negative $\beta$ the stability domain is increased. The stabilization (destabilization) of negative (positive) Hall $\beta$ is a very common phenomenon of all the models. In other words, positive $B_\phi$ (i.e. $\beta>0$) lead to smaller critical field amplitudes than negative  $B_\phi$ (i.e. $\beta<0$).  Hence, if indeed the nonaxisymmetric Tayler instability limits the strength  of the induced toroidal fields $B_\phi$  then the resulting amplitudes are different for different signs of $B_\phi$ due to the action of the Hall effect.
\subsection{Cell structure}
\begin{figure}[htb]
\includegraphics[scale=0.45]{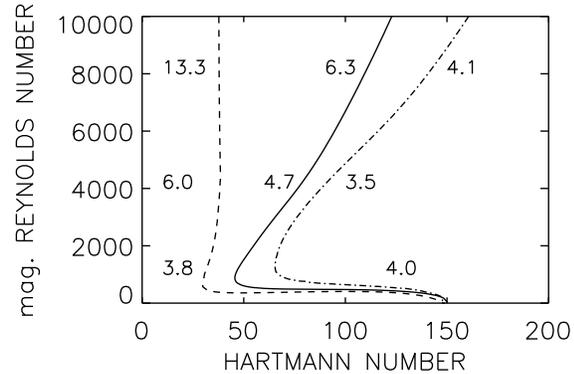}
\caption{The same as in Fig. \ref{mainplots} (bottom, $\rm Pm =100$ ). Stability only exists at left from the curves. They  are marked with the wavenumbers of the marginal instability. The numbers show that also the cell structure strongly depends on the sign of the Hall effect. }
\label{wavenumbers}
\end{figure}

The cell structure of the neutrally stable modes is represented by the resulting vertical wavenumber $k$. From the normalizations it follows the relation
\beg
\frac{\delta z}{R_{\rm out}-R_{\rm in}} = \frac{\pi}{k} \sqrt{\frac{\hat \eta}{1-\hat\eta}}
\label{delta1}
\ende
for the vertical cell size in units of the gap width so that for $\hat \eta=0.5$
\beg
\frac{\delta z}{R_{\rm out}-R_{\rm in}} = \frac{\pi}{k}.
\label{delta2}
\ende
 Hence, for $k\simeq \pi$ the cells are spherical while for $k \gg \pi$ they are rather flat. Both possibilities are realized in the calculations. In Fig. \ref{wavenumbers} the results for $\rm Pm=100$ are extended to much higher values of the magnetic Reynolds number. The difference of the stability domains for different Hall parameters grows even bigger as the rotation increases. The curves are marked with the corresponding wave number values. We find that for negative $\beta$ the cells are spherical but they are rather flat for positive $\beta$. Thus, not only the stability domains strongly differ for the Tayler instability for opposite signs of the  Hall parameter but also the shape of the nonaxisymmetric Tayler vortices depends on that sign. If indeed realized in nature then the {\em sign of the toroidal magnetic field} (in relation to the rotation axis) can easily be read from the observations.
\section{Growth rates}
\begin{figure}[htb]
\mbox
{\includegraphics[scale=0.23]{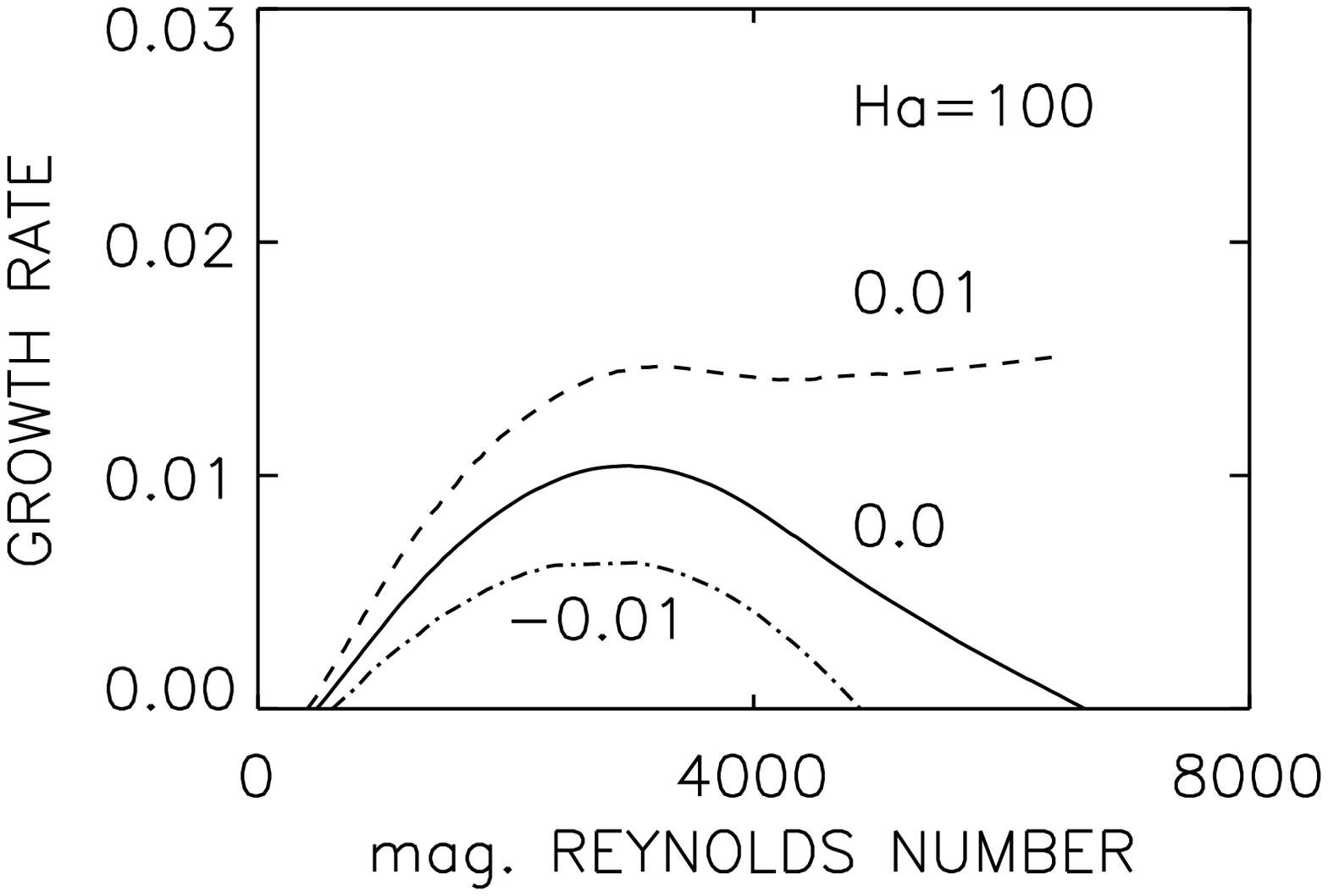}
\includegraphics[scale=0.23]{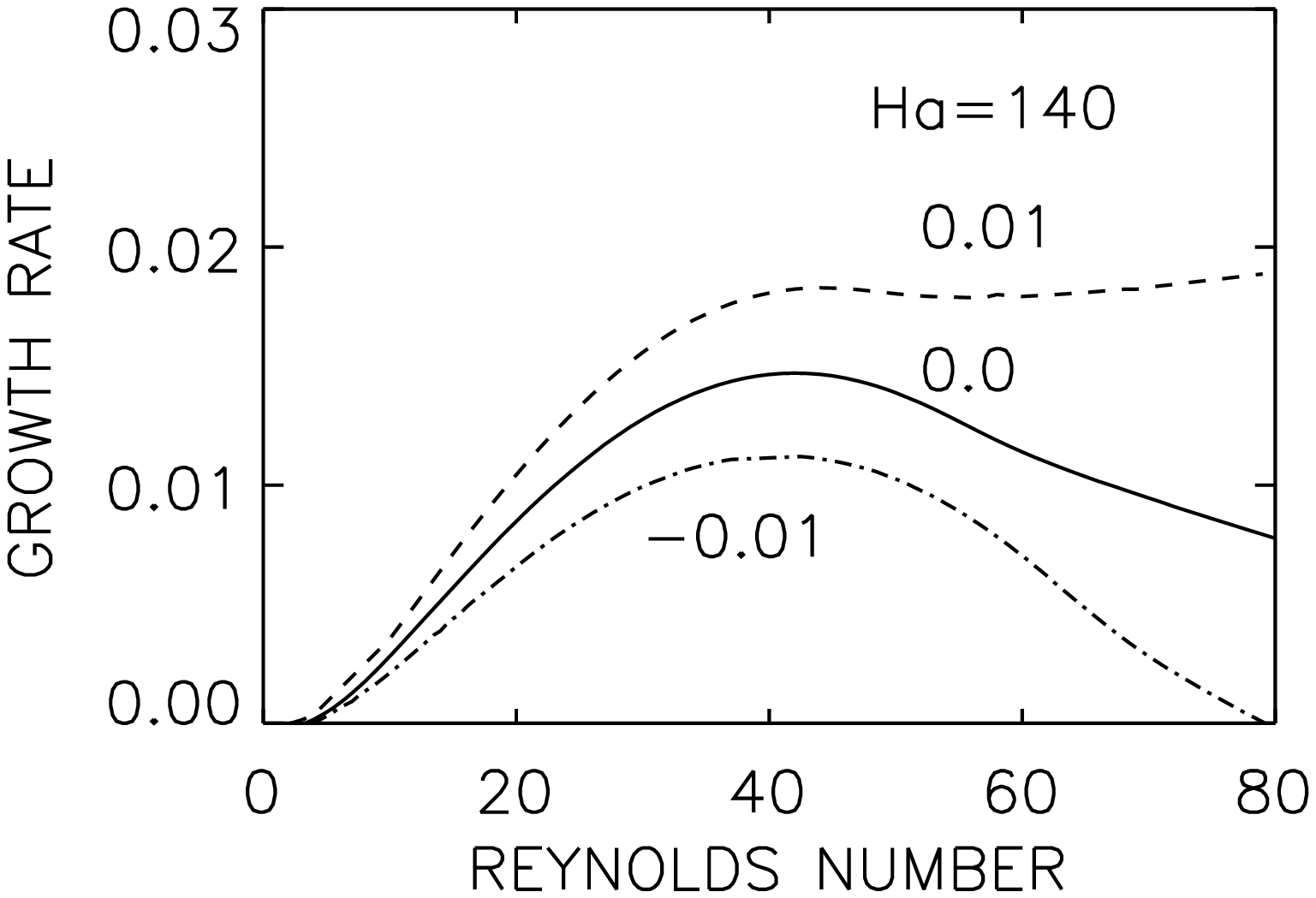}}
\caption{The growth rates of the Tayler instability  normalized with $\Omega_{\rm in}$ for $m=1$ with and without Hall effect. The curves are marked by their Hall parameter $\beta_0$. The values belong to a vertical line for $\rm Ha=100$ (left) and  $\rm Ha=140$ (right) in Fig. \ref{wavenumbers}. $\rm Pm =100$, $\mu_B=1$, $\mu_\Omega=0.5$.}
\label{growth}
\end{figure}

As the Hall time is much longer than the rotation time the question arises whether the Hall effect strongly enhances the growth times of the Tayler instability. However, for the considered parameters the Hall effect  hardly influences the growth rates of the Tayler instability (Fig. \ref{growth}).  Nevertheless, 
the Hall effect  plays an important role  for the stability map of the Tayler instability but not in the resulting linear growth rates of the unstable disturbances. The growth rates are computed along  vertical lines at $\rm Ha=100$  and $\rm Ha=140$  in Fig. \ref{wavenumbers}. The curves are marked with the Hall parameter $\beta_0$ also including $\beta=0$. The growth rates are given in units of the angular velocity of the inner cylinder, at the stability lines they  vanish. A growth rate of 0.01  means an e-folding time of the instability of about 16 rotation periods. For $\rm Pm=100
$ this is the characteristic value for the Tayler instablity without Hall effect. This time is decreased by positive Hall effect and it is increased by negative Hall effect. For positive Hall effect the Tayler instability results as much faster than the Tayler instability for negative Hall effect. All the growth rates grow with growing Hartmann numbers.

Even a weak Hall effect does not generally prolong the growth time of the Tayler instability which  scales with the rotation time. In this case the Hall effect is only a modification of another instability. Even if the Hall effect itself forms the instability (together with the differential rotation) also then the (`shear-Hall') instability scales with the rotation rate and not  with the rather long Hall time (R\"udiger \& Kitchatinov 2005). 

Another example for this phenomenon is given by the plane-wave solution of an $\alpha-\Omega$ dynamo. Both growth rate and cycle time of the most unstable mode of a linear oscillating $\alpha\Omega$ dynamo with weak $\alpha$-effect are mainly fixed by the basic rotation: $\gamma/\Omega \propto (\omega_\eta/ \Omega)^{1/3}$. Here $\gamma$ is the growth rate, $\omega_\eta$  the dissipation frequency and $\Omega$ the basic rotation.

\section{Steeper magnetic profile}
The radial profile of the toroidal magnetic field used in Fig. \ref{mainplots} is rather smooth. Without detailed simulations one cannot  not know the real profile. Hence, the computations represented in Fig \ref{mainplots} are thus repeated for different magnetic field profiles for the most interesting case of high magnetic Prandtl number ($\rm Pm=100$).

\begin{figure}[h]
\includegraphics[scale=0.47]{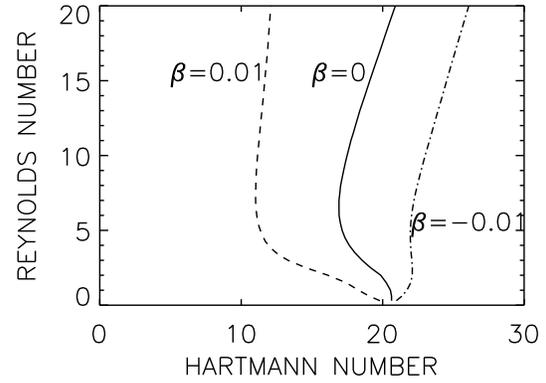}
\caption{Tayler instability ($m=1$) for steep magnetic field ($\mu_B=3$), for  large magnetic Prandtl number
($\rm Pm = 100)$ and with Hall effect.  $\mu_\Omega=0.5$. The curves are labeled by their Hall parameter $\beta_0$.}
\label{highmub}
\end{figure}

Figure \ref{highmub} has been obtained for  magnetic fields that increase outwards ($\mu_B=3$). According to the Tayler criterion (\ref{tay}) such profiles are highly destabilizing. Consequently, we find the critical Hartmann number for $\rm Re=0$ one order of magnitude smaller than in Fig. \ref{mainplots}. The opening of the two curves for $\beta=\pm 0.01$ is with about factor 2 for $\rm Rm\simeq 1000$ very similar to the previous case. Again, generally the stability domain for positive $\beta$ is much smaller than for negative $\beta$. These basic findings do not depend on the actual Hartmann numbers for various magnetic profiles. Nevertheless, we should underline that with the given parameters ($\rho\simeq 10^{13} {\rm g/cm^3}, \nu\simeq 10^9 {\rm cm^2/s}, \eta\simeq 10^7 {\rm cm^2/s}$) for $\rm Ha\simeq 100$ the maximal stable magnetic field is $3\times 10^{12}$ G which value even grows for thinner layers.

\section{Asymmetry of the neutron star hemispheres}
Wardle (1999) has  shown that due to the Hall effect the stability properties of
a  differentially rotating MHD flow depend on the sign of the axial magnetic field.
After our  results  the same is true for the azimuthal magnetic field.
Moreover, the critical magnetic field value above which the flow becomes unstable
can basically differ  for different   magnetic field orientations (Fig. \ref{mainplots}, bottom).

If the effect is strong enough this finding can have  consequences. If in a PNS with differential rotation the toroidal field results from a poloidal field with dipolar symmetry  then also the $B_\phi$ is antisymmetric with respect to the equator.  If the Tayler instability indeed determines the maximal field amplitudes then due to the Hall effect the  amplitudes  of the toroidal field in both hemispheres become different. Obviously, the Tayler-Hall  instability  produces an extra quadrupolar component of the originally produced toroidal fields with dipolar symmetry. It is thus unavoidable that the amplitudes of the induced toroidal field belts are different in both hemispheres.

For $\rm Ha \sim 100$, $\rm Pm \sim 100$ and $\beta_0 \sim 10^{-2}$ taken from the bottom plot
of Fig. \ref{mainplots} we find for the Hall parameter $\hat\beta \sim 10$  leading to  $\sim 10^{13}$ G
for the neutron star. This value is  typical for  pulsars so that the
conclusion about different toroidal field values in both the  hemispheres
of a neutron star due to  Tayler-Hall effects becomes realistic.

On the other hand, strong magnetic fields suppress the heat transport in neutron stars (Schaaf 1988, 1990; Heyl \& Hernquist 1998). The heat transport is blocked in the direction perpendicular to the field
lines so that the heat conductivity tensor becomes anisotropic, i.e.
\beg
\chi_{ij}=\chi_1 \delta_{ij} + \chi_2 B_i B_j,
\label{chi1}
\ende
where $\chi_1$ represents the heat flux perpendicular to the field which is quenched by strong magnetic fields, hence (say) $\chi_1\propto 1/(1+\hat\beta^2)$. With
\beg
\chi_{ij}=\frac{\chi_0}{1+\hat\beta^2} (\delta_{ij} + \hat\beta^2 \frac{B_i B_j}{B^2})
\label{chi2}
\ende
the heat flux remains finite along the field lines even for $B \to \infty$.

The consequence of this magnetic-induced anisotropy of the heat flux tensor is a global inhomogeneity of the surface temperature  (Geppert et al. 2006). If the latitudinal distribution of the magnetic field is strictly symmetric or antisymmetric with respect to the equator then the surface temperature results as equatorsymmetric.
 This is not  true if for both hemispheres the magnetic amplitudes are differing (or, in other words, if the total
 magnetic field
 is a combination of a dipole and a quadrupole). Exactly this is the case if the toroidal magnetic field is produced by  differential rotation under the presence of Tayler-Hall  instability. If the differential rotation of  the neutron star disappears then the magnetic fields are frozen in so that  the magnetic constellation is conserved  for the time scales of the Ohmic decay (also modified by the Hall effect). We  do thus  expect the two half spheres of an isolated neutron star to be of different X-ray activity.

 Schwope et al. (2005) have indeed found an equatorial-asymmetric X-ray brightness  analyzing XMM observations of the isolated neutron star RBS1223. The authors have assumed the existence of one bright ``spot" in each of the hemispheres and found two temperature maxima of different strength (ratio $\epsilon = 0.91$). If this asymmetry effect is general for neutron stars then the interior magnetic fields must also be asymmetric with respect to the equator (dipole plus quadrupole) which can be explained with the Tayler-Hall scenario with differential rotation developed in the present paper.

\acknowledgements
D.A.S. acknowledges the financial support from the
  Deutsche Forschungsgemeinschaft. The simulations were performed with
  the computer cluster SANSSOUCI of the AIP.

\end{document}